# Bi-Filar Coil Winding for Fast Quench Protection

Steven T. Krave and Vittorio Marinozzi

1 *Abstract*— As superconducting magnet technology is pushed towards higher performance, energy density and total stored energy follow exponentially. Protecting magnets becomes substantially more challenging with traditional methods being stretched to their limits. New technologies such as CLIQ (Coupling Loss Induced Quench) promise to provide a robust method to protect advanced magnets, however they become inductance limited in large magnet strings or at low field, leading to more complex configurations. A technique to substantially reduce this limitation and improve response time is presented, by winding coils in a bifilar fashion and connecting them in series for typical operation, while providing an anti-parallel connection for quasi-zero-inductance in a protection case. This allows for extremely high di/dt. The concept is then demonstrated on a small REBCO coil.

## I. Introduction

HISTORICALLY superconducting magnets have been fabricated from coils wound from individual conductors [1]. Some work has been presented using the technique of "two in hand" winding, or using multiple parallel conductors to reduce conductor length, or effective coil inductance, with some implementations to utilize the nth conductor for energy extraction [2, 3, 4]. In one case parallel powered windings with a center tap were explored for a low inductance method to induce rapid quench back [5].This method is similar in that it a secondary current is applied in a non-inductive fashion, however the coils are operated in a series mode, guaranteeing uniform current density and reducing required current leads by 1. This paper details the technique of using a multi-filar approach to coil winding, along with bussing techniques to allow for substantial, generally unexplored degrees of freedom in magnet design. One potential application of this freedom is explored, including electrical models on large accelerator magnet coils, as well as a small high temperature superconducting coil model as a proof of principle. Finally, additional applications are discussed.

## II. The Bifilar Concept

In the case of a bifilar wound magnet coil, two independently insulated cables are wound concurrently. With multiple conductors occupying nearly the same space, they are very tightly magnetically coupled with high mutual inductance, with coupling coefficient (K) values of 0.9 or higher being reasonable to achieve, depending on overall geometry. A coil wound with two conductors allows for a handful of general configurations with a single power supply. They may be generalized as sets of series/parallel inductors with close coupling.

- Single coil powered
- Series powered (Additive)
- Series powered (Differential)
- Parallel powered (Additive)
- Parallel powered (Differential)

While any configuration may find use in certain applications, for the purpose of the non-inductive mode presented here, we are interested in two: The series additive mode and the differential parallel mode. In the series additive mode, the amp-turns add and generate a magnetic field in the normal fashion. The system inductance in the configuration is equal to the standard coil inductance, and magnetic field is generated as normal. In the differential parallel mode, power can be applied across both windings, with one coil amp-turns cancelling the other without generating a substantial net magnetic field. This is essentially a dead short, with some minimal inductance contributed by leakage inductance in the system. This configuration is in many ways the same as would be seen in a resistive superconducting fault current limiter with an additional transport current [6]. A cartoon of the co-winding configurations can be seen below in Fig. *1* A more detailed electrical configuration can be found in Fig. *2*.

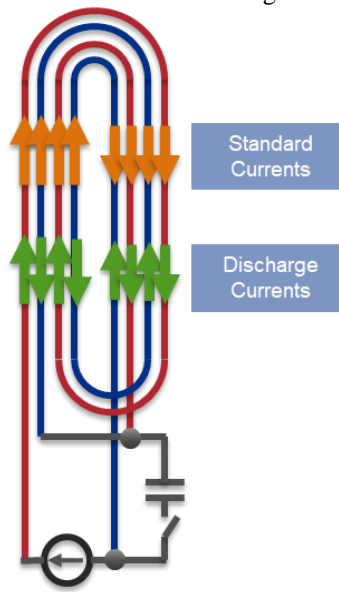

**Fig. 1.** Generalized Bifilar coil layout with capacitor circuit.

The second configuration allows a superposition of cancelled currents on top of the magnet transport configuration without an inductive penalty. In other words, it becomes possibly to drive substantial di/dt in one coil, provided that the opposite condition is met in the other(s) with or without the present of a typical transport current. To drive this di/dt, a "small" voltage is required. This voltage may come from an external source such as shown as a capacitor circuit above, or from interrupting the coil transport current in one loop, while allowing bypass, such as through a diode, in the other. If provided using an external capacitor bank, a third power lead needs to be brought to the joint between coils, as is done with CLIQ.

### III. APPLICATION TO QUENCH PROTECTION

Quench protection for superconducting magnets can be a challenging subject, balancing system response with fragile components. Recently a system to use coupling losses (CLIQ) to generate heat in a coil to cause it to transition to the normal state has been used with excellent success in the Hi-Lumi Quadrupoles and other magnets [7] [8] [9]. The system consists of a capacitor discharged across magnet coils, leading to coupling losses from the large induced *di/dt* and minor corresponding *dB/dt*. CLIQ is an excellent technique to begin exploration of protecting a bifilarly wound coil, as it exploits a high rate of change of current. This configuration fits nicely with the bifilar wound coil. The electrical schematic can be seen below in Fig. *2*. Note that this is the same electrical configuration as the traditional CLIQ, with the addition of an intentional coupling between coils.

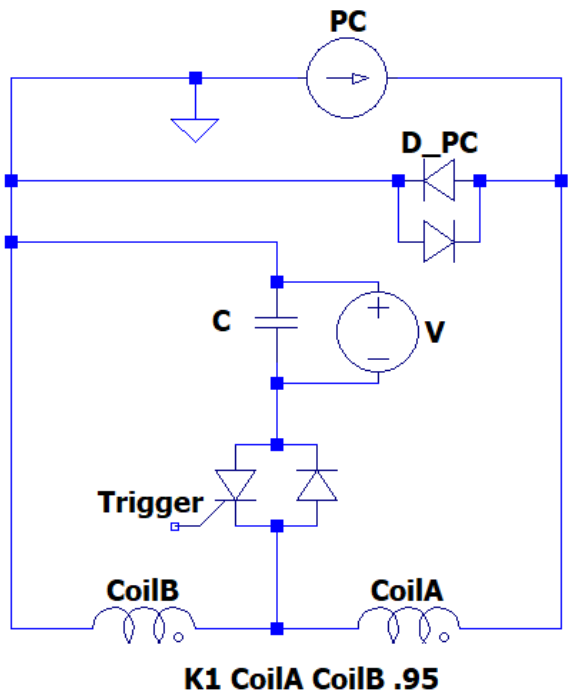

**Fig. 2.** Equivalent circuit approximation. Note that this is nearly electrically the same as CLIQ, with the addition of a high coupling coefficient between coils. The switch shown is triggered by the protection system.

In normal operation, operation current flows around the outside of the loop generating field as normal. In protection mode the capacitor induces counterflowing current in the coils with any needed power convertor current passing through the diodes. Since inductance is near zero, in familiar configurations, di/dt of tens of mega-amps per second can be produced, resulting in massive transport current increase possibly exceeding the short sample limit, as well as introducing substantial AC losses.

For reference, the peak di/dt can be calculated simply by solving (1) below where *V* is capacitor charge voltage and *L* is the differential inductance.

$$V = L * di/dt \qquad (1)$$

The peak current flowing from the capacitor in the ideal case, neglecting lead resistances and dynamic effects, can be estimated by conservation of energy by solving for the capacitor current *I* below, where C is capacitance in Farads, V is charge voltage and L is the differential mode inductance in Henries. The peak current in either coil is ½ of the capacitor current as it is split between coils.

$$\tfrac{1}{2}*C*V^2 = \tfrac{1}{2}*L*I^2 \qquad (2)$$

While theoretically the concept is simple, there are some complications that should be considered or perhaps exploited. A minimum of one additional power lead is required to be connected to the magnet from an external location, like as is required in CLIQ, that is required to pass a massive transport current, albeit for a very short period. This has the additional drawback of an added heat load to the cryostat. This lead needs to be sized to pass the discharge current without substantially impacting the energy available to the magnet or overheating, while minimizing the associated heat leak to an acceptable level. Temperature rise can be calculated adiabatically in the same fashion as the hot-spot temperature. [10]

Additionally, the large additional transport current provided to the magnet will generate some additional forces within the coil. The bulk change in force of the magnet is near zero as ampere turns in the magnet remain approximately constant, as any positive current in one turn is equal, but opposite current in the adjacent turn. There are attractive or repulsive Lorentz forces at individual conductors proportional to *dI*×*B* where *dI* is change in transport current for an individual conductor and *B* is the magnetic field. The outcome of this force is uncertain. It may be beneficial to relax any internal stresses as was shown with the quench current boosting device or may lead to degradation from mechanical damage. [11].

Unfortunately, as presented, this system primarily utilizes over-current to quench a magnet, which requires some operating field/current to ensure that the critical surface is within reach, which may reduce the effectiveness at low current, as is seen with CLIQ, albeit for differing reasons. Additional manipulations may be available to partially address these weaknesses to be discussed in future work.

## IV. QUENCH SIMULATION

We conducted quench simulations to estimate the quench delay time following the discharge of a capacitor in a bifilar coil made from state-of-the-art $Nb_3Sn$ conductor. Specifically, we opted for the LHC Hi-Lumi MQXF conductor, reducing the number of strands from 40 to 28 [10]. We designed a simple 1-meter-long cosine-theta coil using ROXIE, consisting of 32 turns (as shown in the design in **Fig.** *3*). [11, 12]

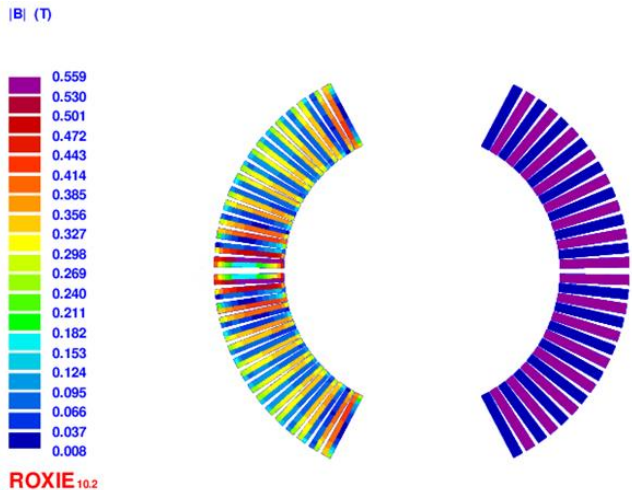

**Fig. 3.** Coil used for the simulations. The left side of the image shows field change at peak discharge current, the right side shows current direction. The purple and blue conductors represent which coil a conductor is a member of, as noted in Fig. 2. Purple cables are a member of coil A, and blue for coil B. When the capacitor is discharged into the system, coil B receives the 1st overcurrent.

The coil's inductance measures 0.2 mH. To simulate the current discharge, we employed the program STEAM-LEDET [13, 14, 15]. This software was chosen for its suitability in simulating inter-filament and inter-strand coupling currents, which play a significant role in inducing quench events due to the high di/dt, in addition to currents surpassing the critical threshold. Furthermore, this software has already undergone experimental validation for the selected conductor [16].

We simulated the discharge of a 40 mF, 1 kV capacitor. This capacitor is compatible with the existing CLIQ units used for HL-LHC triplet magnet protection. The results indicate that it is possible to induce extremely rapid current oscillations, even in accelerator-like superconducting magnets. Indeed, as shown in **Fig.** *4*, the current oscillations in the two halves of the coil exhibit a frequency of approximately 2 kHz, significantly faster than typical CLIQ-induced current oscillations (~50 Hz) manipulating only the coupling coefficient. While the high frequency alone is not claimed to improve performance, any hysteresis and coupling losses may be manipulated to increase power dissipated within the magnet during discharge. This allows for short sample to be exceeded in both coils in the same polarity if desired. While this method does not produce a drastic change in field (the flux is indeed almost constant, being the differential inductance almost zero), roughly 0.25 T evenly over the entirety of the coil, the speed at which it operates leads to the potential for very high dB/dt and therefore high coupling losses [7]. Peak dB/dt in this model was 7.2 times higher than the baseline CLIQ configuration on the same coil. The impacts of this dB/dt need additional study. Moreover, the current briefly multiplies by a factor of at least 4, momentarily exceeding the superconductor's critical current and consequently initiating a quench. This process, however, does not generate an excessive amount of heat.

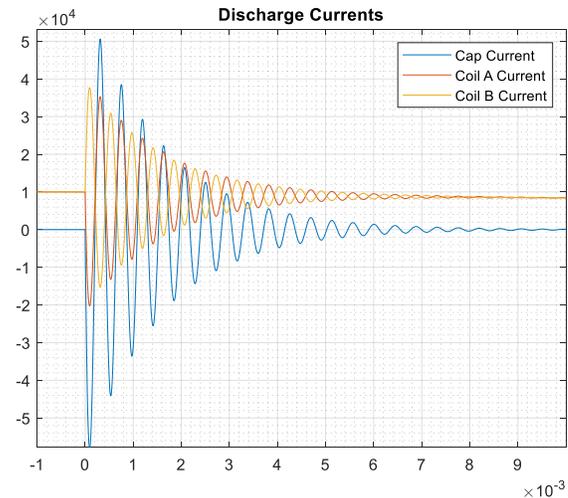

**Fig. 4** Fast current oscillations in the coil shown in Fig 3. The entirety of the discharge energy has dissipated in the magnet within 10 ms.

The combination of high current and the AC losses induced by the high di/dt and corresponding dB/dt can induce a quench on the microsecond scale, as demonstrated in Fig. *5*.

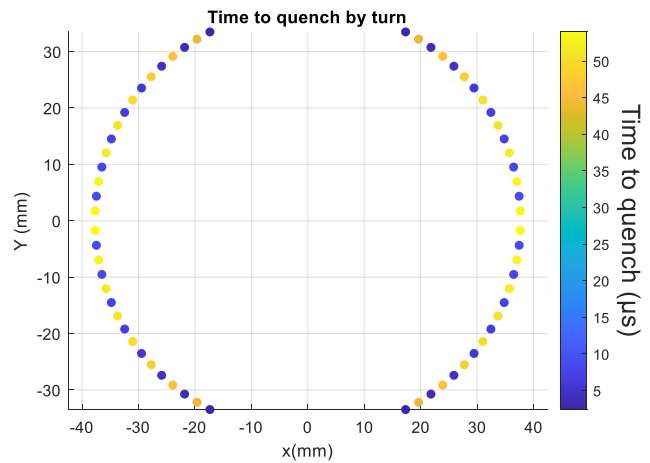

**Fig. 5**. Quench time for each turn of the simulated bifilar coil. Note that the protection system is triggered at 0 s, such that half of all turns quenched at 8 µs with the remainder quenched by 50 µs.

State-of-the-art quench protection systems, such as CLIQ and quench heaters, achieve similar results (complete quenching of the coil) but on a much slower timescale, typically 10 milliseconds or more [17]. Thus, this quench protection method holds significant promise. In the future, we will develop dedicated simulation tools to estimate quench protection results for real-scale, existing accelerator magnets. Our current limitations are primarily due to software constraints, restricting us to small coils at present.

## V. REBCO COIL DESIGN AND TEST

To validate that magnetic coupling between superconducting coils works at least moderately like any other magnetically coupled system, a small magnet was designed and fabricated with a pair of co-wound solenoid coils of 10mm inner radius and 10 turns each, 20 turns total. Parts were 3D printed in ABS or machined from OFHC Copper. The completed coil is shown in **Fig. 6**.

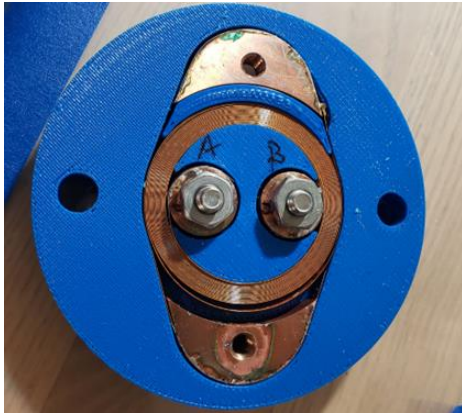

**Fig. 6.** Small Bifilar Solenoid. Not visible on backside is copper bus tying the inner lead from coil A to the outer lead of coil B. Transport current is applied from the outer lead of coil A to the inner lead of coil B.

Individual coil inductances were measured and compared to the series and parallel inductances which closely matched analytical calculations for similar geometry in round wire. Coil inductances as well as other relevant information can be seen below in Table 1. The overall coupling coefficient in this small model coil was reasonable at 0.97 based on the remaining (leakage) inductance in the anti-parallel mode, likely because of the relatively large loop area of the leads with respect to the size of the coil. The mutual inductances between half coils is reported as $M_{bifi}$. Multiple powering configurations were used with an anticipated peak di/dt of 300 MA/s based on measured coil properties.

TABLE 1
COIL DESIGN PARAMETERS

| Geometry Parameters | |
|---|---|
| N Turns (total) | 20 |
| Inner Radius [mm] | 10 |
| $I_{SS}$, 0T, 77K [A] | 95 |
| Outer Ideal Radius [mm] | 14 |
| Total Conductor Length [m] | 1.51 |
| **Magnetic Parameters** | |
| $L_{halfcoil}$ (meas, avg 1k-500k) [H] | 3.7E-06 |
| $L_{series}$ (meas, avg 1k-500k) [H] | 13.9E-06 |
| $L_{antiparallel}$ (meas, avg 1k-500k) [H] | 196E-09 |
| K (Coupling Coeff) | 0.97 |
| $M_{Bifi}$ [H] | 3.58E-06 |
| **Powering parameters** | |
| $V_{charge}$ [V] | 5 to 60 |
| C [μF] | 50 to 40000 |
| di/dt @ 0v | 300E6 Max |

Several powering configurations were used; first, a low capacitance film capacitor of ~400 μF to demonstrate the fast, non-inductive response, and second, a larger 40mF cap to allow enough time to measure coil resistance. Both configurations showed initial di/dt matches the analytical calculation, however it is difficult to resolve a resistance change in the high frequency case. In all cases a 1 mΩ shunt resistor was used in line with the capacitor, with an additional shunt resistor used to measure the transport current. A large choke (1 mH) was installed in line to the power supply to prevent deviation from the set steady state current.

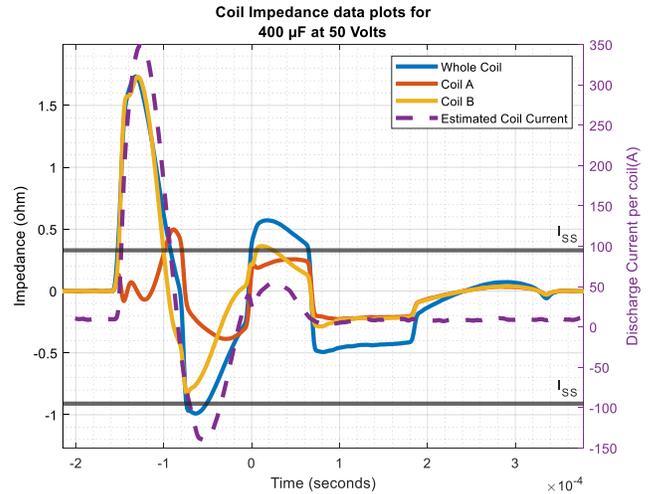

**Fig. 7.** Discharge into coil with 400 μF capacitor for oscillation frequency of ~12 kHz. Note that coil impedance as defined in the plots is the coil voltage over the normal transport current. This results in the half coil impedance values showing an impact of the the discharge which is cancelled in the whole coil (bucked) signal.

While some impedance can be measured as $V/I_{norm}$ where $I_{norm}$ is the normal transport current (3 A), the parasitic inductance and capacitance in the system make it difficult to interpret the system behavior, especially including the additional voltage drop from the diode pack. It is clear that we are able to transfer energy to the system at ~22 kW, much of this goes into ESR, thyristor voltage drop, and capacitor lead voltage drop. At high currents(>100A), the diodes used (70HFR120) have an effective contribution to the whole coil of <10 mΩ and are considered negligible. The available power after these losses at the coil peaks at ~2.5 kW. This test case developed a peak boost current of +/- 350 A in 25 μs. Note that the analytical calculation shows higher boost currents, indicating some clamping effect on the system.

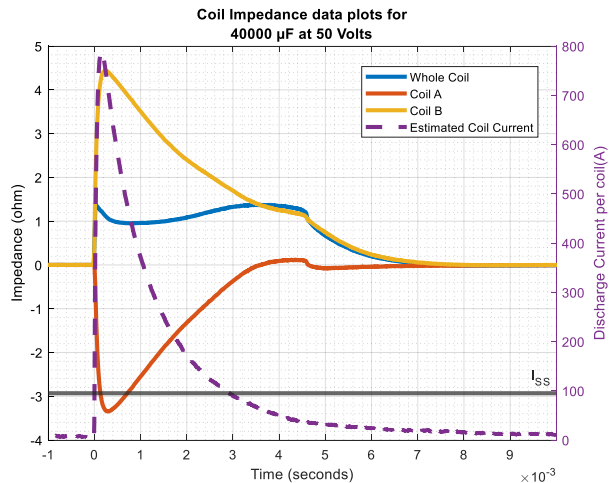

**Fig. 8.** Discharge using 40 mF capacitor with no oscillation. A coil resistance of ~1Ω is developed in a few microseconds and is maintained until the coil recovers below short sample limit in a few ms.

In the case with no oscillations and a large capacitor, the initial inductive behavior of the system holds with high di/dt, and peak discharge power of ~50 kW. Without oscillation, a larger fraction of the stored energy, around 40%, is transferred to the coils. This test case developed a peak boost current of +/- 800 A in 150 μs with respect to the conductor short sample limit of 96 A as measured at 77 K, 0 T. In this case, the analytical boost current is +/- 4.5 kA per coil, indicating strong clamping in the coil as tested. In both cases the coil recovered quickly.

In general, the testing shows that non-inductive behavior is achievable through bifilar winding and boost currents can be reasonably predicted analytically. This should be expected as similar coil architectures are used frequently for other superconductor applications.

## VI. POTENTIAL ADDITIONAL APPLICATIONS

Having additional co-wound conductors in a magnet enables several improvements on existing technology. These are to be further explored in the future. Some straightforward options are listed below:

### A. Transport Current Manipulation for Quench Avoidance

The high coupling between co-wound coils allows near instantaneous transfer of current from one loop to another. In the case of the onset of a superconducting transition, it may be possible to redirect some current from one coil to another, through the use of multiple power supplies or active current control via superconducting switches or power electronics. The simulation done earlier in LEDET shows that this may be reasonable as the second coil avoids quench until the local SSL is exceeded.

### B. Quench Detection

A bifilar wound coil allows near perfect magnetic coupling of parallel windings. This technique is frequently applied for voltage taps to reduce any inductive pickup in differential sensing. This would provide many of the benefits found with co-wound voltage taps, without needing to co-wind delicate additional wires. [18] [19] A true bifilar coil has the potential reduce measurement noise substantially.

### C. Increase of Coil Current Density

A reduction in quench detection time and protection time enabled by the above provides substantial overhead in magnet protection with respect to hotspot temperature and reduces the demands for additional copper in the wire cross section [18]. This provides a clear path towards substantial engineering current density ($J_e$) increases by simply reducing the copper present in the strand provided that the conductor remains stable.

### D. Differential Tuning

Having a second near identical set of conductors enables ultraprecise tuning of harmonics at the expense of slightly reduced coupling with the addition of a second tuner power supply circuit. Additional co-wound coils add degrees of freedom of tuning parameters. Sinusoidally modulated coils such as the canted cosine theta design with slightly different coefficients per harmonic are a simple example.

## VII. CONCLUSIONS

Multi-filar coils for superconducting magnets are an enabling technology based on existing technologies with well understood behavior. Quench protection techniques based on capacitor discharge as explored here are seen as one way to drastically improve magnet performance. We observed that it is possible to induce a quench through the discharge of a capacitor in a bifilar coil. However, the understanding of the entity of the quench and of its capability to protect large magnets will require a whole magnet design philosophy to be used to reach substantially higher performance. Additional magnet optimization routes become available for implementation.

## REFERENCES

[1] M. N. Wilson, "Superconducting Magnets," Oxford University Press, 1983, p. 307.
[2] M. Mentink, A. Dudarev, T. Mulder, J. Van Nugteren and H. ten Kate, "Quench Protection of Very Large, 50-GJ-Class, and High-Temperature-Superconductor-Based Detector Magnets," in IEEE Transactions on Applied Superconductivity, vol. 26, no. 4, pp. 1-8, June 2016, Art no. 4500608, doi: 10.1109/TASC.2015.2510078.
[3] N. Tesla, "Coil for Electro-magnets". United States Patent US512340A, 7 July 1893.
[4] M. A. Green, "Can High Current Density HTS Magnets be Quench Protected Using Methods Used to Protect High Current Density LTS Magnets?," in IEEE Transactions on Applied Superconductivity, vol. 31, no. 5, pp. 1-6, Aug. 2021), Art no. 4701006, doi: 10.1109/TASC.2021.3058203.
[5] T. Wakuda, Y. Ichiki and M. Park, "A Novel Quench Protection Technique for HTS Coils," IEEE Transactions on Applied Superconductivity, vol. 22, no. 3, pp. 4703404-4703404, 2011.
[6] M. Noe and M. Steurer, "High-temperature superconductor fault current limiters: concepts, applications, and development status," Superconductor Science and Technology, vol. 20, no. 3, p. R15, 2007.
[7] E. Ravaioli et al., "Modeling of Interfilament Coupling Currents and Their Effect on Magnet Quench Protection," in IEEE Transactions on Applied Superconductivity, vol. 27, no. 4, pp. 1-8, June 2017, Art no. 4000508, doi: 10.1109/TASC.2016.2636452.[8] G. Ambrosio and et al, "MQXFA Final Design Report," US HL-LHC Accelerator Upgrade Project, 2022.
[9] E. Ravaioli,. (2015). CLIQ A new quench protection technology for superconducting magnets. 10.3990/1.9789036539081.
[10] L. D. Cooley, A. K. Ghosh, D. R. Dietderich and I. Pong, "Conductor Specification and Validation for High-Luminosity LHC Quadrupole Magnets," in IEEE Transactions on Applied Superconductivity, vol. 27, no.

4, pp. 1-5, June 2017, Art no. 6000505, doi: 10.1109/TASC.2017.2648738.
[11] N. Schwerg, H. Henke and S. Russenschuck, "Numerical calculation of transient field effects in quenching superconducting magnets," 2009.
[12] S. Russenschuck, "Field Computation for Accelerator Magnets: Analytical and Numerical Methods for Electromagnetic Design and Optimization," Weinheim, Germany: Wiley-VCH, 2010, pp. 575-608.
[13] CERN, "STEAM-LEDET," [Online]. Available: https://espace.cern.ch/steam/_layouts/15/start.aspx#/LEDET/Home.aspx.
[14] E. Ravaioli, B. Auchmann, M. Maciejewski, H.H.J. ten Kate, A.P. Verweij, "Lumped-Element Dynamic Electro-Thermal model of a superconducting magnet," Cryogenics, Volume 80, Part 3, 2016, Pages 346-356, ISSN 0011-2275, https://doi.org/10.1016/j.cryogenics.2016.04.004.
[15] L. Bortot et al, "STEAM: A hierarchical co-simulation framework for superconducting accelerator magnet circuit," IEEE Transactions on Applied Superconductivity, vol. 28, no. 3, pp. 1-6, April 2018.
[16] E. Ravaioli et al., "Quench Protection Performance Measurements in the First MQXF Magnet Models," in IEEE Transactions on Applied Superconductivity, vol. 28, no. 3, pp. 1-6, April 2018, Art no. 4701606, doi: 10.1109/TASC.2018.2793900.
[17] S. Izquierdo Bermudez et al., "Overview of the Quench Heater Performance for MQXF, the Nb3Sn Low-β Quadrupole for the High Luminosity LHC," in IEEE Transactions on Applied Superconductivity, vol. 28, no. 4, pp. 1-6, June 2018, Art no. 4008406, doi: 10.1109/TASC.2018.2802839.
[18] L. Bottura, "Magnet Quench 101," in Workshop on Accelerator Magnets, Superconductor, Design and Optimization, 2013.
[19] N. Bykovskiy, D. Uglietti, P. Bruzzone and K. Sedlak, "Co-Wound Superconducting Wire for Quench Detection in Fusion Magnets," in IEEE Transactions on Applied Superconductivity, vol. 32, no. 4, pp. 1-5, June 2022, Art no. 2500105, doi: 10.1109/TASC.2022.3140706.